\documentclass[12pt,a4paper]{article}

\usepackage{lineno,hyperref}
\modulolinenumbers[5]

\usepackage{amsmath,amssymb}
\usepackage{bm}
\usepackage{graphicx}
\usepackage{subcaption}
\usepackage{natbib}
\usepackage{authblk}
	
\title{Non-identifiability of parameters for a class of shear-thinning rheological models, with implications for haematological fluid dynamics}

\author{M.T.\ Gallagher$^{1,2,3}$, R.A.J.\ Wain$^{4,5,6}$, S.\ Dari$^{2}$\\ J.P.\ Whitty$^5$, D.J.\ Smith$^{2,3}$}

\affil{\small{$^1$m.t.gallagher@bham.ac.uk\\
	   $^2$School of Mathematics, $^3$Institute for Metabolism and Systems Research, and $^4$Institute of Translational Medicine, University of Birmingham, B15 2TT, UK\\
	   $^5$John Tyndall Institute, School of Engineering, and $^6$School of Medicine and Dentistry, University of Central Lancashire, Preston, PR1 2HE, UK}}
\date{}
	
\begin{document}

\maketitle

\begin{abstract}
	Choosing a suitable model and determining its associated parameters from fitting to experimental data is fundamental for many problems in biomechanics. Models of shear-thinning complex fluids, dating from the work of Bird, Carreau, Cross and Yasuda, have been applied in highly-cited computational studies of heamodynamics for several decades. In this manuscript we revisit these models, first to highlight a degree of uncertainty in the naming conventions in the literature, but more importantly to address the problem of inferring model parameters by fitting rheology experiments. By refitting published data, and also by simulation, we find large, flat regions in likelihood surfaces that yield families of parameter sets which fit the data equally well. Despite having almost indistinguishable fits to experimental data these varying parameter sets can predict very different flow profiles, and as such these parameters cannot be used to draw conclusions about physical properties of the fluids, such as zero-shear viscosity or relaxation time of the fluid, or indeed flow behaviours. We verify that these features are not a consequence of the experimental data sets through simulations; by sampling points from the rheological models and adding a small amount of noise we create a synthetic data set which reveals that the problem of parameter identifiability is intrinsic to these models.
\end{abstract}

\section{Introduction}
Many complex fluids exhibit shear rate-dependent viscosity; suspensions, in particular fluids of biological importance such as blood, and biological polymers, such as mucus, are typically shear-thinning (pseudo-plastic), i.e.\ their viscosity reduces with increasing shear rate. A number of models have been proposed for this behaviour and studied intensively; we will focus on a class of models which relate the shear viscosity to shear rate via nonlinear algebraic equations, in particular the formulations of Cross, Bird, Carreau and Yasuda, and their subsequent application to blood rheology. We will address two significant issues -- first, an inconsistency in the literature regarding naming of models, and more importantly, some significant difficulties which appear in determining model parameters through least squares fitting. Since there are major (and unexpected) differences in parameter identifiability between subtly different models, unambiguous naming will turn out to be very important. To set the scene we will briefly review the key models.

The earliest model of \cite{ostwald1925} and \cite{deWaele1923} is based on a power-law dependence of viscosity on shear rate; limitations of this simple model include is its singularity at zero shear rate and inability to capture high shear rate dependency when compared to empirical data. For these reasons we will not consider the model further. These discrepancies were addressed by \cite{cross1965} who postulated a four parameter, constitutive relationship:
\begin{equation}
\mu(\dot{\gamma}) = \mu_\infty + \frac{\mu_0-\mu_\infty}{1+(\lambda \dot{\gamma})^{1-n}}, \label{eq:cross}
\end{equation}
where $\mu$ is the effective viscosity of the fluid as a function of shear rate \(\dot{\gamma}\), the parameters $\mu_0$ and $\mu_\infty$ are the zero and infinite limit shear viscosities respectively, $\lambda$ is a constant with dimensions of time, and $n$ is the power-law index (equation \eqref{eq:cross} is presented in a slightly different form from \cite{cross1965} though is functionally identical). The model has finite, non-zero viscosity, at both zero and infinite shear rate limits. By contrast, the three parameter model of \cite{carreau1972} provides a finite viscosity at zero shear rate, and zero viscosity in the infinite shear rate limit,
\begin{equation}
\mu \left( \dot{\gamma} \right) = 
\frac{\mu_0}{\left( 1+\left( \lambda \dot{\gamma} \right)^2 \right)^\frac{1-n}{2}}.
\label{eq:carreau1972}
\end{equation}
The original paper uses the parameter $S=(1-n)/2$. In a study of polystyrene fluids, \cite{yasuda1979} modified this formulation to include a further parameter \(a\) to describe better the low shear to power-law transition region:
\begin{equation}
\mu \left( \dot{\gamma} \right) = 
\frac{\mu_0}{\left( 1+\left( \lambda \dot{\gamma} \right)^a \right)^\frac{1-n}{a}}.
\label{eq:yasuda1979}
\end{equation} 
This model has four free parameters (\(\mu_0, \lambda, n, a\)), and implies a zero viscosity limit as shear rate tends to infinity.

Perhaps surprisingly, the canonical text of \cite{bird1987} 
(while citing the same sources as above) defines a different model as the {\em `Carreau-Yasuda'} model
\begin{equation}
\mu \left( \dot{\gamma} \right) = 
 \mu_\infty + \frac{\mu_\infty- \mu_0}{\left( 1+\left( \lambda \dot{\gamma} \right)^a \right)^\frac{1-n}{a}}.
\label{eq:bird1987}
\end{equation}
Equation~\eqref{eq:bird1987} differs from both Carreau's and Yasuda's models through the inclusion of an infinite shear rate viscosity parameter \(\mu_\infty\) (in the manner of \cite{cross1965}), amounting to five parameters. 

Bird et al.'s five-parameter ``Carreau-Yasuda'' model \eqref{eq:bird1987} has been used for blood flow modelling in key papers by \cite{perktold1995}, \cite{gijsen1999} (who referred to Bird et al.\ and also termed it Carreau-Yasuda) and \cite{leuprecht2001} (who referred to it as a {\em modified Cross} model). Equation~\eqref{eq:bird1987} can be viewed as a hybrid of Carreau, Yasuda and Cross' contributions, which perhaps explains the proliferation of terminology. Indeed for suitably-chosen parameters, equation~\eqref{eq:bird1987} can be reduced to each of the preceding models.

The variability in terminology can also be found in major commercial computational fluid dynamics codes such as ANSYS-Fluent, ANSYS-CFX, and Abaqus (see Table~\ref{tab:cfd}). For the avoidance of confusion, below we refer to equations \eqref{eq:cross}, \eqref{eq:carreau1972},  \eqref{eq:yasuda1979} and \eqref{eq:bird1987} as the Cross-1965, Carreau-1972, Yasuda-1979 and BCCY-1987 (Bird-Cross-Carreau-Yasuda) models respectively.

\begin{table}[htp]
	\centering
	\caption{Three commercial CFD packages and their varying terminology}
	\label{tab:cfd}
	\begin{tabular}{c c c}
		Software & Their name & Equation\\
		\hline 
		Abaqus 		 & Carreau-Yasuda 	& Eq~\eqref{eq:bird1987}\\
		 	   		 & Carreau 			& Eq~\eqref{eq:bird1987} with $ a = 2 $\\
		 	   		 & Cross 				& Eq~\eqref{eq:cross}\\
		ANSYS-CFX 	 & Carreau-Yasuda 	& Eq~\eqref{eq:bird1987}\\
					 & Bird-Carreau		& Eq~\eqref{eq:bird1987} with $ a = 2 $\\
					 & Cross 			& Eq~\eqref{eq:crossZero}\\
		ANSYS-Fluent & Carreau			& Eq~\eqref{eq:bird1987} with $ a = 2 $ and a\\ 
		& & leading temperature factor\\
				 & Cross 			& Eq~\eqref{eq:crossZero} with a leading\\
		& & temperature factor
	\end{tabular}
\end{table}

\section{Data fitting}
Given a functional form \(\mu(\dot{\gamma};\bm{\theta})\) for the viscosity \(\mu\) at shear rate \((\dot{\gamma})\) with parameters \(\bm{\theta}=(\mu_0,n,\ldots)\), and rheometry data \((\dot{\gamma}_m,\mu_m)\), maximum likelihood estimation with the assumption of normally distributed error leads to the least squares parameter estimate,
\begin{equation}
\bm{\theta}^*=\underset{\bm{\theta}}{\mathrm{argmin}}\ \mathcal{L}(\bm{\theta}):= \underset{\bm{\theta}}{\mathrm{argmin}}\sum_{m=1}^M (\mu(\dot{\gamma}_m;\bm{\theta})-\mu_m)^2. \label{eq:mle}
\end{equation}

We will also find it useful to optimize over subsets of parameters while the remaining are held fixed. To denote this we will use, for example, the notation
\begin{equation}
\mathcal{L}(\mu_0,\lambda,*):=\underset{n}{\mathrm{max}}\ \mathcal{L}(\mu_0,\lambda,n).
\end{equation}

Most rheological literature takes the slightly different approach of fitting the model to data on a log-log scale. In the framework of maximum likelihood estimation, this approach corresponds to the assumption of lognormal error. Mathematically, we may define,
\begin{equation}
\bm{\theta}^{\Diamond}=\underset{\bm{\theta}}{\mathrm{argmin}}\ \mathcal{\ell}(\bm{\theta}) := \underset{\bm{\theta}}{\mathrm{argmin}}\sum_{m=1}^M (\log(\mu(\dot{\gamma}_m;\bm{\theta}))-\log(\mu_m))^2.\label{eq:logmle}
\end{equation}

Data from \cite{skalak1981} (extracted from \cite{ballyk1994} using GRABIT, \cite{grabit}) will be used in what follows because of its excellent coverage in shear rate, from \(0.1\)~s\(^{-1}\) to \(500\)~s\(^{-1}\).

\section{Results}
Each model will be considered in turn, starting with the model possessing the fewest free parameters. The units of \(\mu_0,\  \mu_\infty\) will be centipoise, \(\lambda\) will be seconds, and \(n,a\) are dimensionless.

\subsection{Carreau-1972 three-parameter model}
\begin{figure}[tp]
	\centering
	\begin{subfigure}[t]{\textwidth}
		\includegraphics[width=\textwidth]{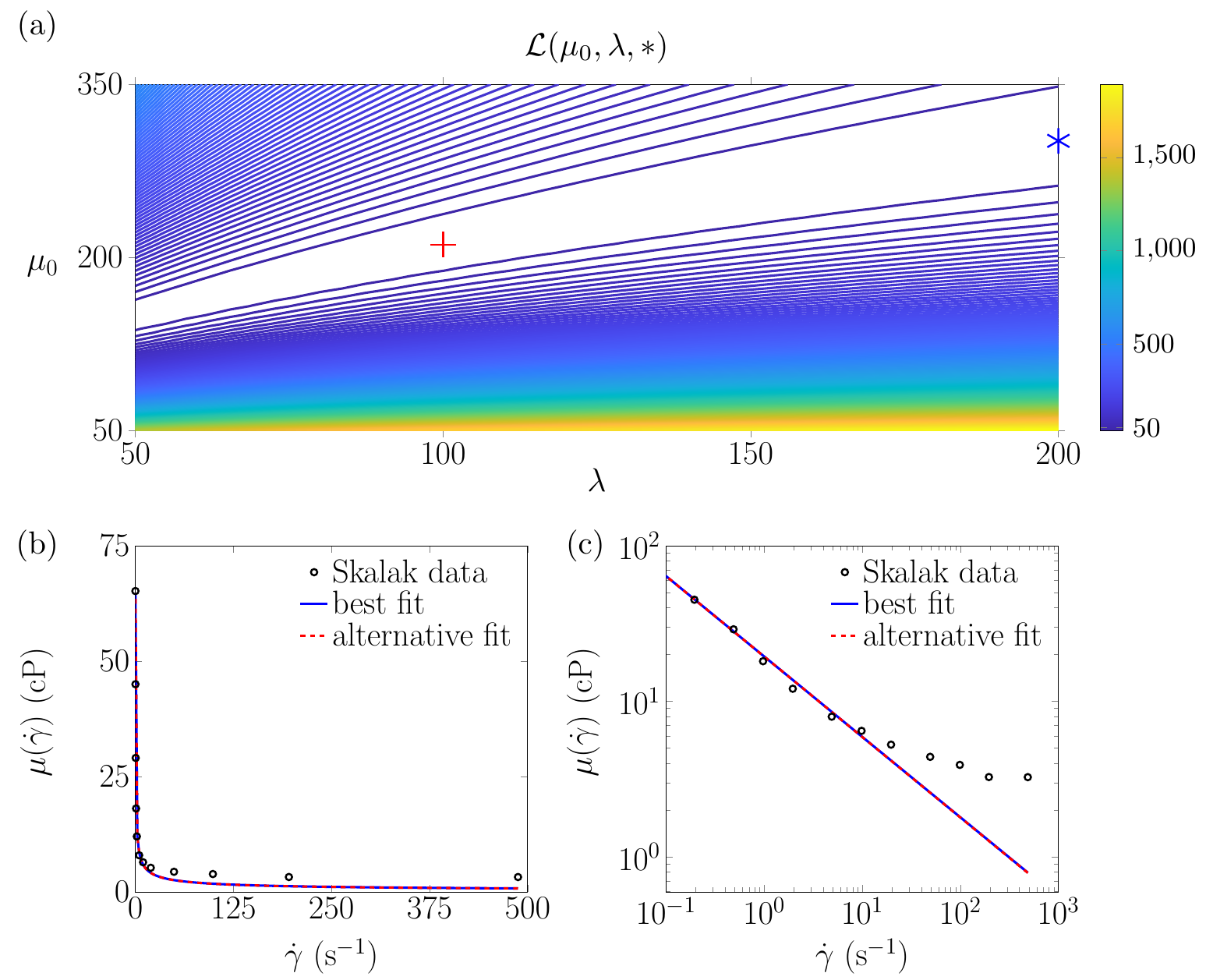}
		\phantomcaption
		\label{fig:carreau1972a}	
	\end{subfigure}
	\begin{subfigure}[t]{0\textwidth}
		\phantomcaption
		\label{fig:carreau1972b}	
	\end{subfigure}
	\begin{subfigure}[t]{0\textwidth}
		\phantomcaption
		\label{fig:carreau1972c}	
	\end{subfigure}
	\caption{Carreau-1972 model fits (normal error) to data of Skalak et al. Panel (a) shows linearly spaced contours of the objective function $ \mathcal{L}\left(\mu_0,\lambda,*\right) $, with a blue star showing the location of the Matlab \texttt{fmincon} parameter fit (with $ n = 0.483$), and a red plus showing the arbitrarily chosen `alternative' parameter choice (with $ n^+ = 0.482$). Panels (b) and (c) plot the fits of the Carreau-1972 model to the data of \cite{skalak1981} for these parameter tuples shown on linear and logarithmic axes respectively.}
	\label{fig:carreau1972}
\end{figure}

Performing a maximum likelihood fit of the Carreau-1972 model with normal errors, i.e.\ by minimizing \(\mathcal{L}\), immediately yields difficulties with parameter identification. Constrained optimization (Matlab \texttt{fmincon}) consistently finds a solution at the boundary of the parameter space (either the maximum value of \(\mu_0\) or the maximum value of \(\lambda\) depending on the limits chosen, however with \(n=0.483\) consistently. The reason for this behaviour is evident in Figure~\ref{fig:carreau1972a}; the likelihood surface \(\mathcal{L}(\mu_0,\lambda,*)\) exhibits an extended `ridge'. Taking an upper bound of \(\mu_0<350\) yields the parameter estimate \(\mu_0=301,\ \lambda=200,\ n=0.483\), depicted with a blue star; the cost function value is \(\mathcal{L}^*=25.8\). To show how indeterminate this fit is, we will examine an arbitrarily chosen `alternative' parameter tuple of \(\mu_0=211,\ \lambda=100\) and fitted optimum \(n^+=0.482\) (red plus), which has a very similar cost function value of \(\mathcal{L}^+=25.9\). For any practical purpose, the fits are identical, as shown in Figure~\ref{fig:carreau1972}b,c. The data do not therefore reliably constrain the parameters \(\mu_0\) or \(\lambda\).
It is also of note that either parameter set fits the data well with these parameters up to approximately \(\dot{\gamma}=10\)~s\(^{-1}\), however they both perform rather badly for higher values of shear rate.

\begin{figure}[tp]
	\centering
	\begin{subfigure}[t]{\textwidth}
		\includegraphics[width=\textwidth]{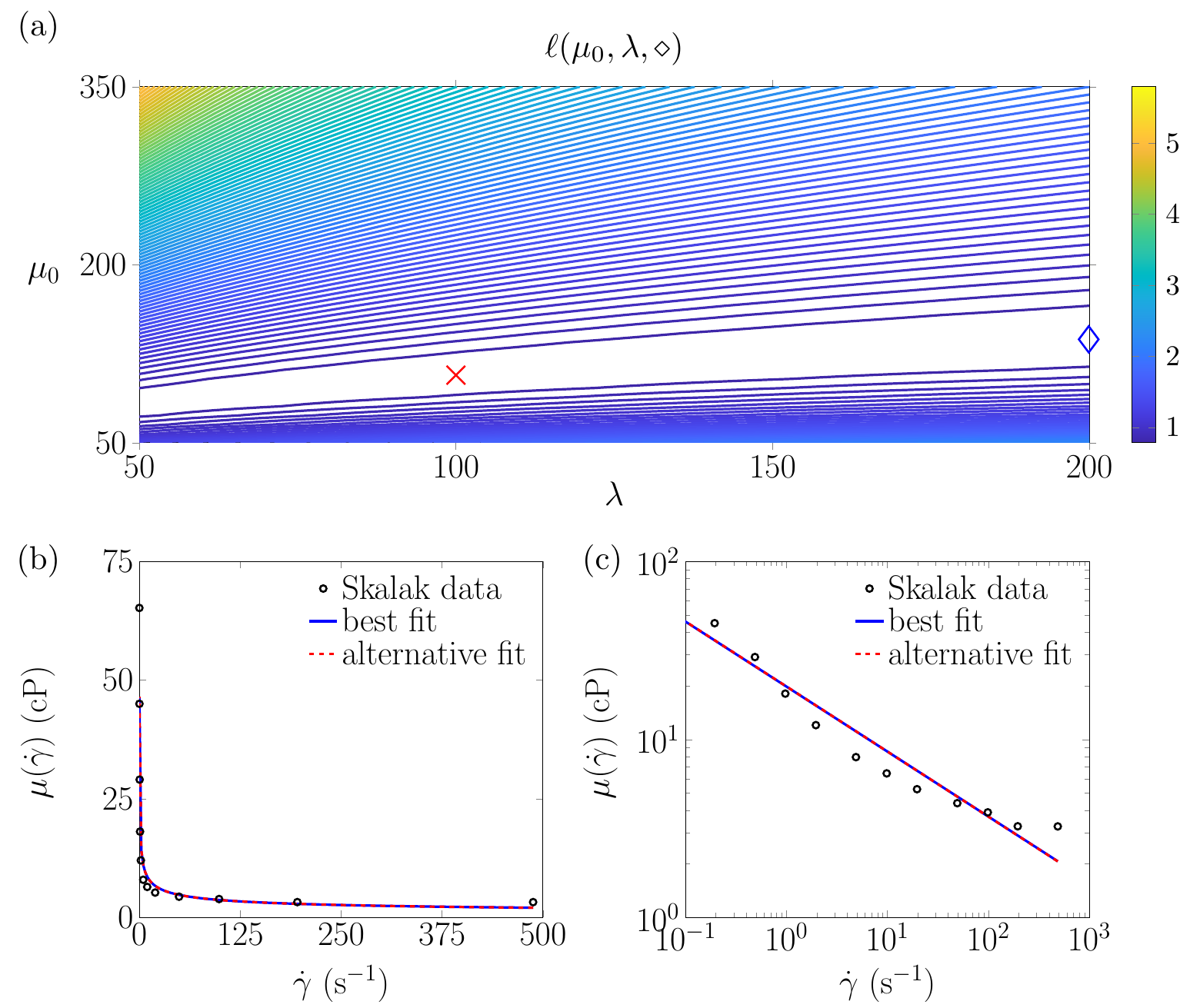}
		\phantomcaption
		\label{fig:logCarreau1972a}	
	\end{subfigure}
	\begin{subfigure}[t]{0\textwidth}
		\phantomcaption
		\label{fig:logCarreau1972b}	
	\end{subfigure}
	\begin{subfigure}[t]{0\textwidth}
		\phantomcaption
		\label{fig:logCarreau1972c}	
	\end{subfigure}
	\caption{Carreau-1972 model fits (lognormal error) to data of Skalak et al. Panel (a) shows linearly spaced contours of the objective function $ \ell\left(\mu_0,\lambda,\diamond\right) $, with a blue diamond showing the location of the Matlab \texttt{fmincon} parameter fit (with $ n^{\diamond} = 0.635$), and a red cross showing the arbitrarily chosen `alternative' parameter choice (with $ n^{\times} = 0.634$). Panels (b) and (c) plot the fits of the Carreau-1972 model to the data of \cite{skalak1981} for these parameter tuples shown on linear and logarithmic axes respectively.}
	\label{fig:logCarreau1972}
\end{figure}

One may ask whether the more traditional approach of fitting to the log-log plot, i.e.\ by taking lognormal error and minimizing \(\ell\), might work better. The results of this process are shown in Figure~\ref{fig:logCarreau1972}. The higher shear rate region (\(10\)--\(500\)~s\(^{-1}\)) is fitted much better, however the indeterminacy issue is still present. The best fit tuple found by \texttt{fmincon} (with the same bounds) is \(\mu_0^\Diamond=137,\ \lambda^\Diamond=200,\ n^\Diamond=0.635\) which has cost function \(\ell^\Diamond=0.732\). A manually and arbitrarily-chosen tuple \(\mu_0^\times=107,\ \lambda^\times=100,\ n^\times = 0.634\), plotted as a red cross, yields a very similar cost function value of \(\ell^\times=0.733\). Again the flow curves corresponding to each parameter set are essentially identical (Figure~\ref{fig:logCarreau1972}a,b). While the fit is arguably better than for Figure~\ref{fig:carreau1972}, the parameters \(\mu_0,\lambda\) are again indeterminate from the data. The same issue occurs for several other experimental blood rheology data sets, see~\ref{sec:appCarreau1972}).

\begin{figure}[tp]
	\centering
	\begin{subfigure}[t]{\textwidth}
		\includegraphics[width=\textwidth]{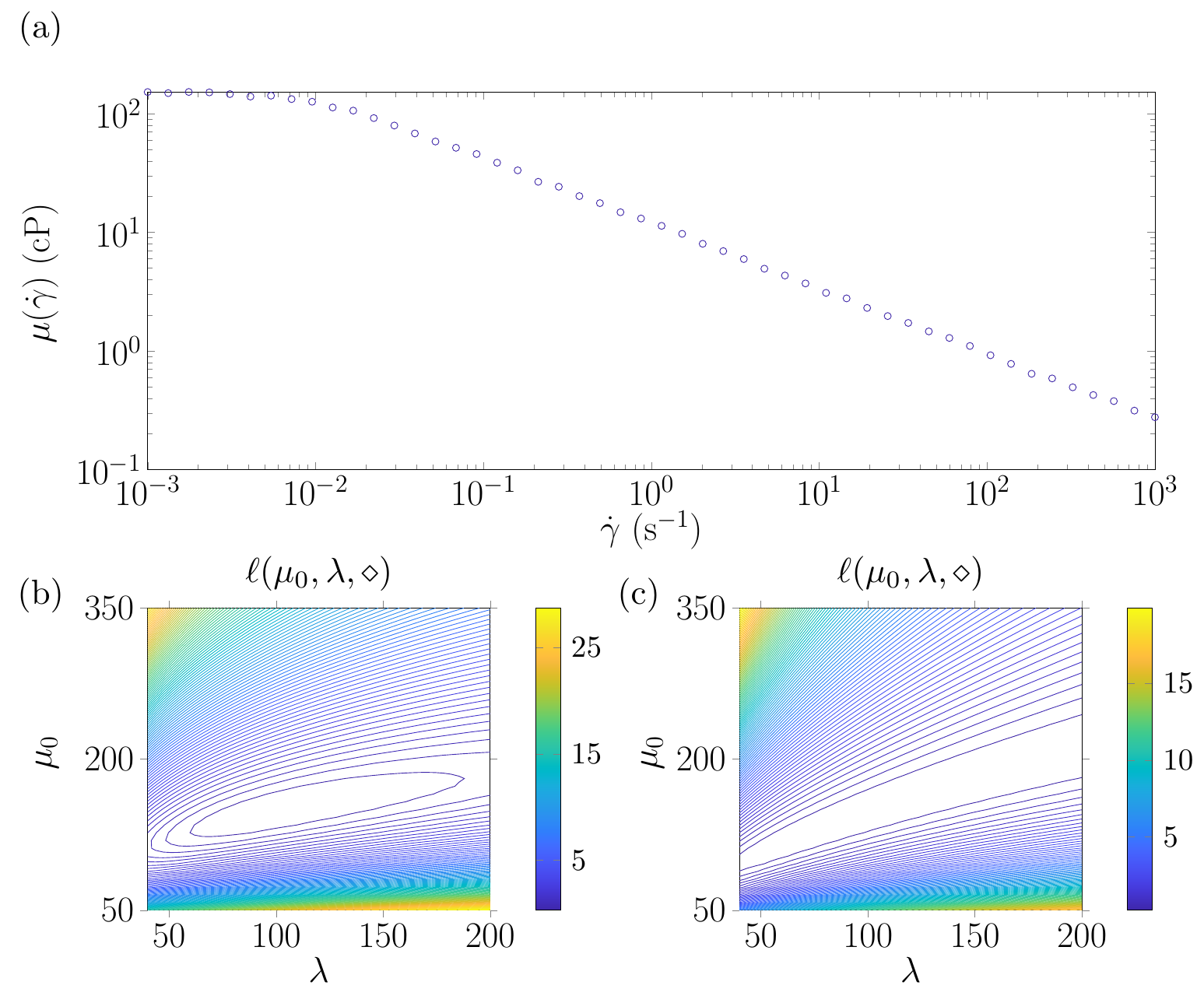}
		\phantomcaption
		\label{fig:synthCarreaua}	
	\end{subfigure}
	\begin{subfigure}[t]{0\textwidth}
		\phantomcaption
		\label{fig:synthCarreaub}	
	\end{subfigure}
	\begin{subfigure}[t]{0\textwidth}
		\phantomcaption
		\label{fig:synthCarreauc}	
	\end{subfigure}
	\caption{Numerical experimental results: synthetic data generated from, then fitted to the Carreau-1972 model. (a) Lognormal noise with standard deviation \(0.02\) was used to generate 50 points of synthetic data from the Carreau-1972 flow model with \(\mu_0=150, \lambda=100, n=0.45\). (b) Cost function, optimized over \(n\) for each \((\mu_0,\lambda)\) tuple for synthetic data generated over the shear rate range (\(10^{-3}\)--\(10^{3}\)~s\(^{-1}\)); (b) Cost function for 50 synthetic data points generated over the shear rate range (\(10^{-2}\)--\(10^{3}\)~s\(^{-1}\)).}
	\label{fig:synthCarreau}
\end{figure}

Is this problem a consequence of the data available, or is it intrinsic to the Carreau-1972 model? We generated a synthetic data set by choosing parameter values and simulating an experimental series of 50 samples, taken over an extremely wide range of shear rates (\(10^{-3}\)--\(10^{3}\)~s\(^{-1}\)) and with a very small addition of lognormal noise (with standard deviation \(0.02\)). The synthetic data are shown in Figure~\ref{fig:synthCarreaua}, with fitting results in Figures~\ref{fig:synthCarreau}b,c. Fitting over the full range in shear rate (Figure~\ref{fig:synthCarreaub}) reveals that a local minimum is now evident, although with again a rather elongated basin. However restricting the range in shear rate at the lower end to \(10^{-2}\)--\(10^{3}\)~s\(^{-1}\) (Figure~\ref{fig:synthCarreauc}), the basin is extended in a similar way to the real data fit of Figure~\ref{fig:logCarreau1972}. While this shows that it is possible to alleviate somewhat the indeterminacy, for blood rheology it requires a range of low shear rate data that may not be possible to achieve in practice. We are not aware of measurements of this type being available in the literature.

\subsection{Cross-1965}
\begin{figure}[tp]
	\centering
	\begin{subfigure}[t]{\textwidth}
		\includegraphics[width=\textwidth]{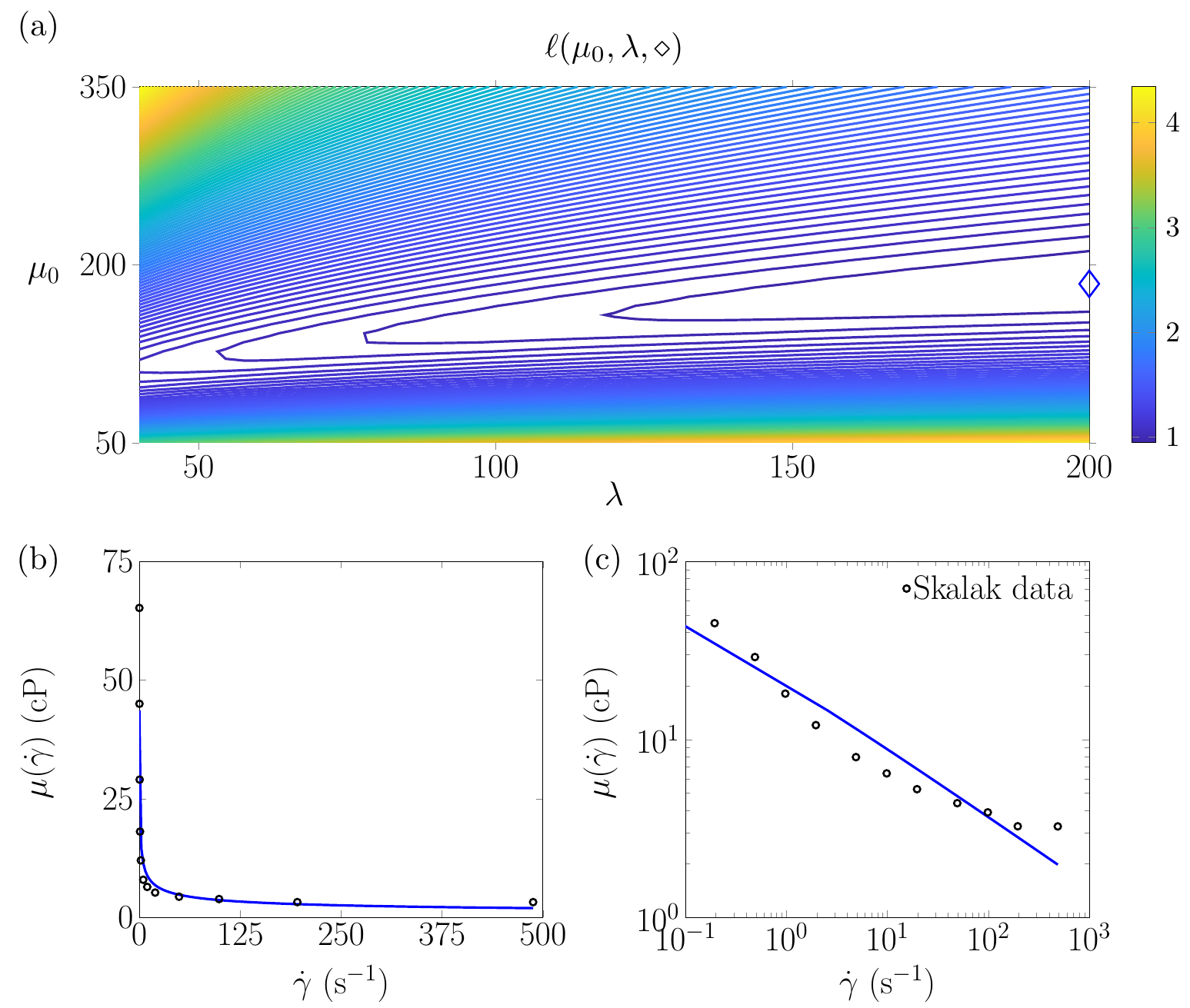}
		\phantomcaption
		\label{fig:crossZeroa}	
	\end{subfigure}
	\begin{subfigure}[t]{0\textwidth}
		\phantomcaption
		\label{fig:crossZerob}	
	\end{subfigure}
	\begin{subfigure}[t]{0\textwidth}
		\phantomcaption
		\label{fig:crossZeroc}	
	\end{subfigure}
	\caption{Results of fitting the Skalak et al.\ data with lognormal error with our `Cross-Zero' model (equation~\eqref{eq:crossZero}, based on Cross-1965 with \(\mu_\infty=0\). Panel (a) shows linearly spaced contours of the objective function $ \ell\left(\mu_0,\lambda,\diamond\right) $, with a blue diamond showing the location of the Matlab \texttt{fmincon} parameter fit. Panels (b) and (c) plot the fits of the Cross-Zero model to the data of \cite{skalak1981} for these parameter tuples shown on linear and logarithmic axes respectively.}
	\label{fig:crossZero}
\end{figure}

As described above, the Cross-1965 model involves the parameters \(\mu_0,\ \lambda,\ n\), and, in addition, an infinite shear rate viscosity \(\mu_\infty\). To facilitate comparison with the Carreau-1972 model we will initially set \(\mu_\infty=0\), which we will refer to as the `Cross-Zero' model,
\begin{equation}
\mu(\dot{\gamma}) =  \frac{\mu_0}{1+(\lambda \dot{\gamma})^{1-n}}. \label{eq:crossZero}
\end{equation}
The result of fitting this model with lognormal error to the data of Skalak et al.\ is shown in Figure~\ref{fig:crossZero}a,b -- while the fit is excellent, the minimum of the cost function again appears at the boundary of the domain. Extending the bounds of the search space has similar effects to the Carreau-1972 model. The reason for this can be seen by inspecting equation~\eqref{eq:crossZero}: for sufficiently large \(\lambda\) and non-zero shear rate, the constitutive law can be approximated by,
\begin{equation}
\mu(\dot{\gamma}) \approx  \frac{\mu_0}{(\lambda \dot{\gamma})^{1-n}}, \label{eq:crossZeroApprox}
\end{equation}
yielding an infinite family of approximately equivalent parameterizations for which \(\mu_0/\lambda^{1-n}\) is constant.

\begin{figure}[tp]
	\centering
	\begin{subfigure}[t]{\textwidth}
		\includegraphics[width=\textwidth]{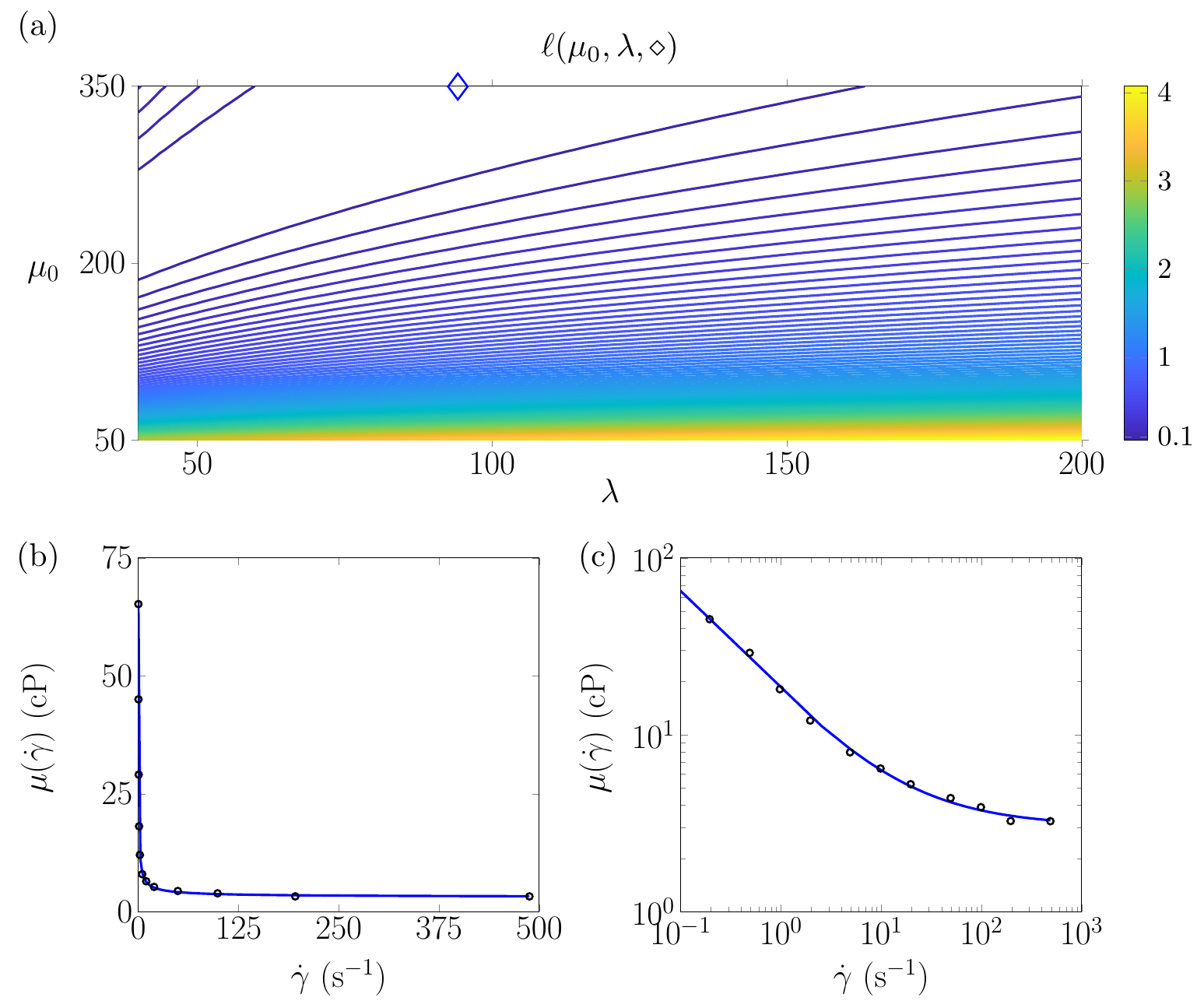}
		\phantomcaption
		\label{fig:crossa}	
	\end{subfigure}
	\begin{subfigure}[t]{0\textwidth}
		\phantomcaption
		\label{fig:crossb}	
	\end{subfigure}
	\begin{subfigure}[t]{0\textwidth}
		\phantomcaption
		\label{fig:crossc}	
	\end{subfigure}
	\caption{Results of fitting the Skalak et al.\ data with lognormal error with the Cross-1965 model (equation~\eqref{eq:cross}). Panel (a) shows linearly spaced contours of the objective function $ \ell\left(\mu_0,\lambda,\diamond\right) $, with a blue diamond showing the location of the Matlab \texttt{fmincon} parameter fit. Panels (b) and (c) plot the fits of the Cross-1965 model to the data of \cite{skalak1981} for these parameter tuples shown on linear and logarithmic axes respectively.}
	\label{fig:cross}
\end{figure}

Having established that this simplified version of the Cross model is also affected by parameter indeterminancy, we turn our attention to the full four-parameter Cross-1965 model (Figure~\ref{fig:cross}). While the fit (Figure~\ref{fig:cross}a,b) is rather better than both the Carreau-1972 and Cross-Zero models, particularly for larger shear rates, a similar parameter indeterminacy occurs as for Carreau-1972 and Cross-Zero (Figure~\ref{fig:crossc}).

\subsection{Yasuda-1979}
\begin{figure}[tp]
	\centering
	\begin{subfigure}[t]{\textwidth}
		\includegraphics[width=\textwidth]{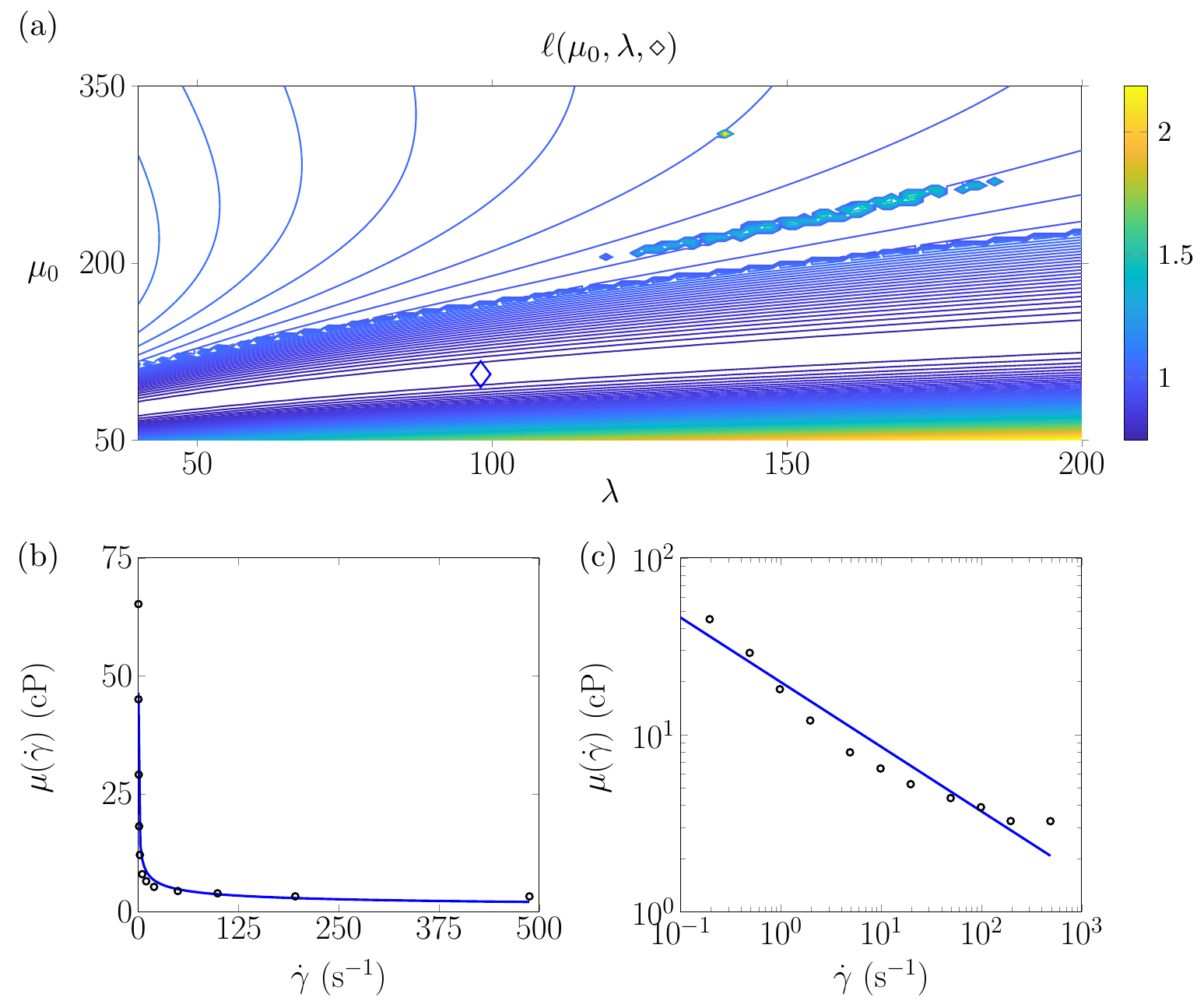}
		\phantomcaption
		\label{fig:yasudaa}	
	\end{subfigure}
	\begin{subfigure}[t]{0\textwidth}
		\phantomcaption
		\label{fig:yasudab}	
	\end{subfigure}
	\begin{subfigure}[t]{0\textwidth}
		\phantomcaption
		\label{fig:yasudac}	
	\end{subfigure}
	\caption{Results of fitting the Skalak et al.\ data with lognormal error with the Yasuda-1979 model (equation~\eqref{eq:yasuda1979}). Panel (a) shows linearly spaced contours of the objective function $ \ell\left(\mu_0,\lambda,\diamond\right) $, with a blue diamond showing the location of the Matlab \texttt{fmincon} parameter fit. Panels (b) and (c) plot the fits of the Cross-Zero model to the data of \cite{skalak1981} for these parameter tuples shown on linear and logarithmic axes respectively.}
	\label{fig:yasuda}
\end{figure}

The Yasuda-1979 model differs from Carreau-1972 only through an additional index parameter. An interesting effect of including this parameter is that a local minimum is now found interior to the search domain (Figure~\ref{fig:yasuda}), at \(\mu_0=106,\ \lambda=98.1,\ n=0.635,\ a=7.61\). Nevertheless, there is still a very elongated ridge in parameter space and associated uncertainty. 

\subsection{BCCY-1987}
\begin{figure}[tp]
	\centering
	\begin{subfigure}[t]{\textwidth}
		\includegraphics[width=\textwidth]{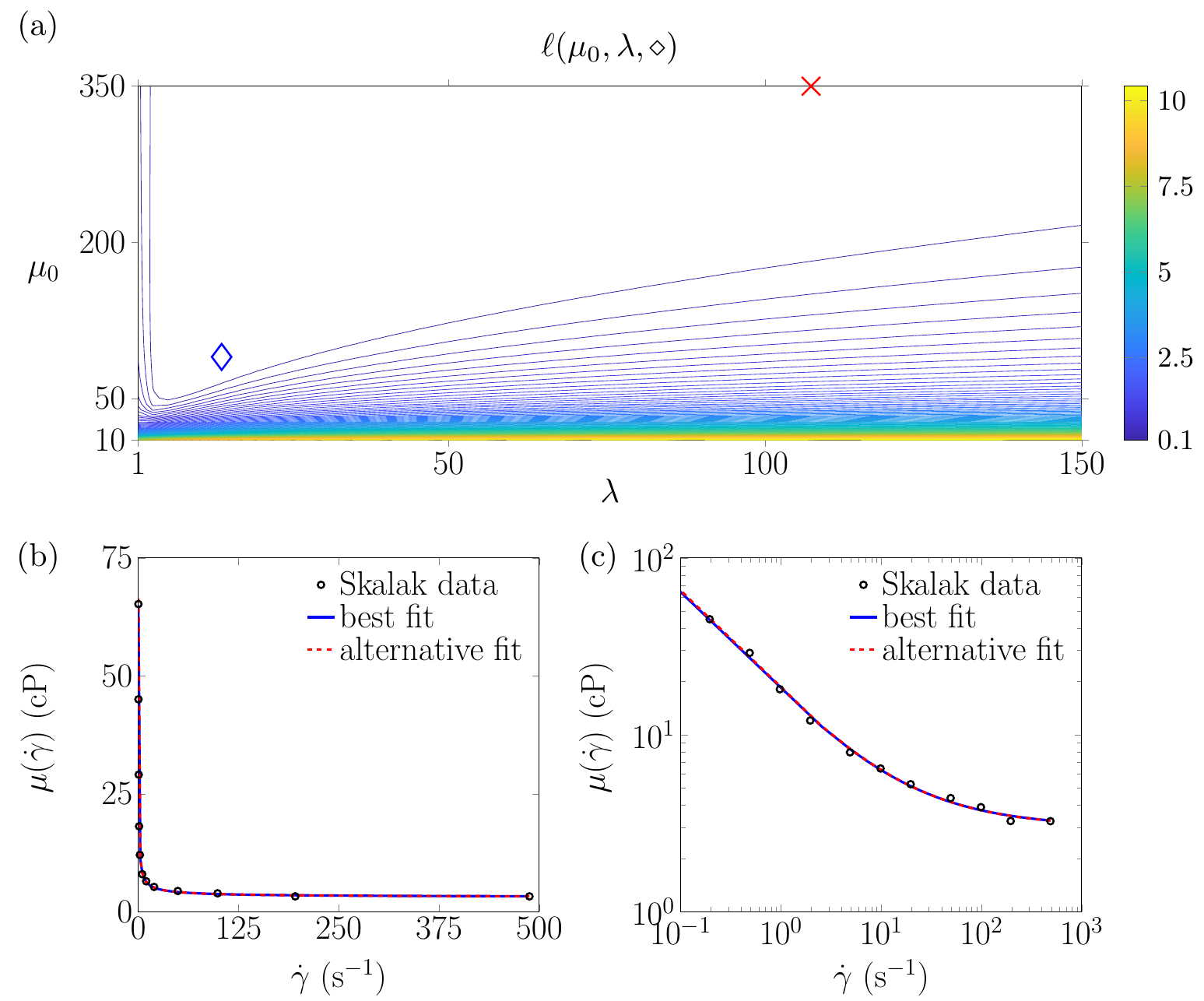}
		\phantomcaption
		\label{fig:BCCYa}	
	\end{subfigure}
	\begin{subfigure}[t]{0\textwidth}
		\phantomcaption
		\label{fig:BCCYb}	
	\end{subfigure}
	\begin{subfigure}[t]{0\textwidth}
		\phantomcaption
		\label{fig:BCCYc}	
	\end{subfigure}
	\caption{Results of fitting the Skalak et al.\ data with lognormal error with the BCCY-1987 model (equation~\eqref{eq:bird1987}). Panel (a) shows linearly spaced contours of the objective function $ \ell\left(\mu_0,\lambda,\diamond\right) $, with a blue diamond showing the location of the Matlab \texttt{fmincon} parameter fit, and a red cross showing the arbitrarily chosen `alternative' parameter choice. Panels (b) and (c) plot the fits of the BCCY-1987 model to the data of \cite{skalak1981} for these parameter tuples shown on linear and logarithmic axes respectively.}
	\label{fig:BCCY}
\end{figure}

Finally, we consider the most general five-parameter BCCY-1987 model (Figure~\ref{fig:BCCY}); the best fit parameter tuple (blue diamond) is \(\mu_0=89.8,\ \mu_\infty=3.03,\ \lambda=14.2,\ n=0.339\ a=2.15\) and the cost function \(\ell^\Diamond=0.0196\). The fit is excellent, as for the Cross-1965 model, and as for the Yasuda-1979 model there is a minimum interior to the domain. However, the indeterminacy is arguably the worst of all models considered, with a large flat region in the top left corner of the parameter domain considered. An alternative point chosen at \(\mu_0=350,\ \lambda=107\) with optimized \(\mu_\infty=3.03,\ n=0.334,\ a=0.755\) yields only a marginal increase in cost function value \(\ell^\times=0.0214\) (red cross). 
 
\section{Discussion}
This paper considered the identification of model parameters from experimental data for various cases of what we have termed the Bird-Cross-Carreau-Yasuda class of steady shear-thinning rheological models, specifically applied to blood data. Given that all of the models considered exhibited significant uncertainty regarding parameter values -- and in the case of the Carreau-1972 and Cross-1965 models the optimum value depended entirely on the specification of the search domain -- it is clear that it is necessary to be cautious regarding the physical interpretation of the parameters derived from such a fit. While the flow index \(n\) was very consistent, the parameters \(\mu_0\) and \(\lambda\) are indeterminate and therefore cannot be used to draw conclusions about the zero shear viscosity or relaxation time of the fluid.

\begin{figure}[tp]
	\centering
	\begin{subfigure}[t]{\textwidth}
		\includegraphics[width=\textwidth]{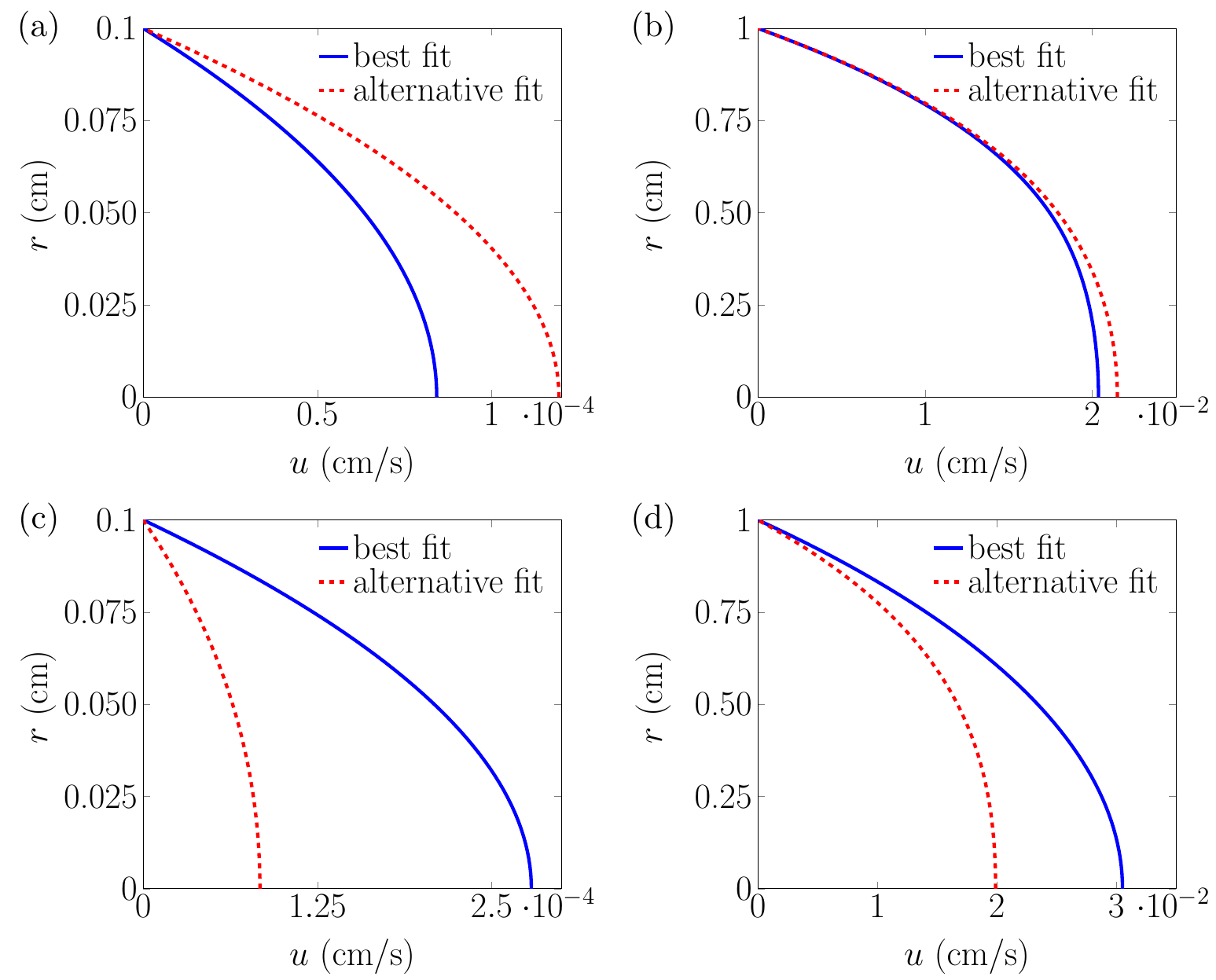}
		\phantomcaption
		\label{fig:pipeflowsa}	
	\end{subfigure}
	\begin{subfigure}[t]{0\textwidth}
		\phantomcaption
		\label{fig:pipeflowsb}	
	\end{subfigure}
	\begin{subfigure}[t]{0\textwidth}
		\phantomcaption
		\label{fig:pipeflowsc}	
	\end{subfigure}
	\begin{subfigure}[t]{0\textwidth}
		\phantomcaption
		\label{fig:pipeflowsd}	
	\end{subfigure}
	\caption{Velocity profiles computed for pipe flow due to pressure gradient \(10\)~dyn/cm for the best and alternative parameter fits for (a,b) Carreau-1972, (c,d) BCCY-1984. (a,c) pipe radius \(0.1\)~cm and (b,d) pipe radius \(1\)~cm.}
	\label{fig:pipeflows}
\end{figure}

One may ask whether this parameter indeterminacy actually matters for flow simulation. After all, if one has an accurate model of the response of the fluid to a range of shear rates, why would the individual values of the parameters used to produce this curve matter? To provide insight into this question, we computed pressure-driven axisymmetric pipe flow with the Carreau-1972 and BCCY-1987 models with each of the `best fit' and `alternative fit' parameter choices. The results are shown in Figure~\ref{fig:pipeflows}. In all cases the pressure gradient was chosen as \(10\)~dyn/cm. For each case there is a significant relative difference between the flow profiles for each parameter fit. Parameter indeterminacy may therefore significantly affect flow predictions, particularly for flows involving low shear rates.

The rheology of shear-thinning fluids, and indeed the specific field of blood rheology, are much wider-ranging than the class of models and steady-shear experiments we have considered here. For recent review, see \cite{anand2017} and examples of recent on time-dependent flows and extensional rheometry, see references \cite{apostolidis2015} and \cite{kolbasov2016}. The importance of the Bird-Cross-Carreau-Yasuda class of models is underscored by the fact that they have formed part of major works such as the highly cited papers of \cite{perktold1995} and \cite{gijsen1999}. As such the matter of parameter identification for these models is important to address. Our investigation has shown that parameter fitting for this class of models is indeterminate for several key blood rheology data sets in the literature. Moreover, this is still an issue even for simulated high accuracy data over a very wide range of parameter sets. The implications of this indeterminacy are that parameter values for \(\mu_0\) and \(\lambda\) in particular cannot be physically interpreted, and predictions from pipe flow models may also be subject to uncertainty. In future studies -- not just involving the BCCY class of models -- it will be important to assess parameter sensitivity in order to have confidence in model predictions.

\section*{Data accessibility}

All data and code for generating the figures in this report can be accessed in the GitLab repository:

 \url{https://gitlab.com/meuriggallagher/nonnewtonianparamident}

\section*{Acknowledgements}

M.T.G. and D.J.S. are supported by the Engineering and Physical Sciences Research Council award EP/N021096/1. S.D. was supported by the London Mathematical Society Undergraduate Research Bursary 17-18 13. This work started with the Multi-scale Biology Study Group, University of Birmingham (12-15th December 2016), which was jointly funded by POEMS (Predictive modelling for healthcare technology through maths - EP/L001101/1) and MSB-Net (UK Multi-Scale Biology Network - BB/M025888/1). The sponsors had no role in the study design.

\section*{Conflict of interest statement}

The authors confirm that there are no financial or personal relationship
with other people or organisations that could inappropriately
influence (bias) this work.

\appendix

\section{Indeterminacy of the Carreau-1972 model applied to other datasets}
\label{sec:appCarreau1972}
\begin{figure}[tp]
	\centering
	\begin{subfigure}[t]{\textwidth}
		\includegraphics[width=\textwidth]{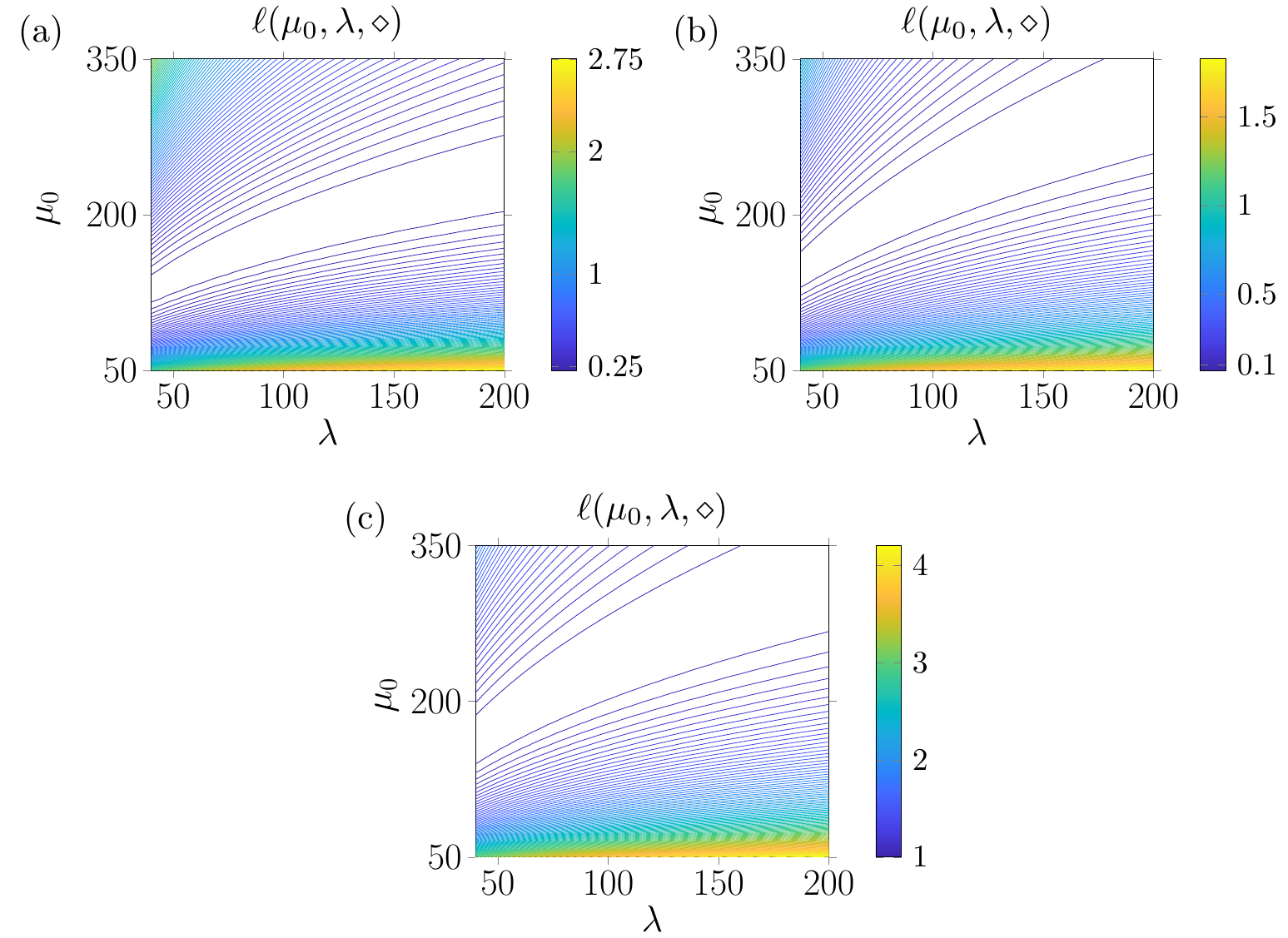}
		\phantomcaption
		\label{fig:appCarreau1972a}	
	\end{subfigure}
	\begin{subfigure}[t]{0\textwidth}
		\phantomcaption
		\label{fig:appCarreau1972b}	
	\end{subfigure}
	\begin{subfigure}[t]{0\textwidth}
		\phantomcaption
		\label{fig:appCarreau1972c}	
	\end{subfigure}
	\caption{Likelihood surfaces optimized over \(n\) for the Carreau-1972 model with lognormal error, consistently showing extended flat regions. (a) Merrill et al., (b) Cokelet et al., (c) Huang et al.}
	\label{fig:appCarreau1972}
\end{figure}

The parameter fitting to the Carreau-1972 model in Figure~\ref{fig:logCarreau1972c} is applied to the combined data sets from \cite{merrill1963}, \cite{cokelet1963} and \cite{huang1973} in Figure~\ref{fig:appCarreau1972}. Each of these data sets have been extracted from \cite{ballyk1994} using GRABIT, \cite{grabit}.

\begin{thebibliography}{18}
\expandafter\ifx\csname natexlab\endcsname\relax\def\natexlab#1{#1}\fi
\providecommand{\url}[1]{\texttt{#1}}
\providecommand{\href}[2]{#2}
\providecommand{\path}[1]{#1}
\providecommand{\DOIprefix}{doi:}
\providecommand{\ArXivprefix}{arXiv:}
\providecommand{\URLprefix}{URL: }
\providecommand{\Pubmedprefix}{pmid:}
\providecommand{\doi}[1]{\href{http://dx.doi.org/#1}{\path{#1}}}
\providecommand{\Pubmed}[1]{\href{pmid:#1}{\path{#1}}}
\providecommand{\bibinfo}[2]{#2}
\ifx\xfnm\relax \def\xfnm[#1]{\unskip,\space#1}\fi
\bibitem[{Anand \& Rajagopal(2017)}]{anand2017}
\bibinfo{author}{Anand, M.}, \& \bibinfo{author}{Rajagopal, K.~R.}
  (\bibinfo{year}{2017}).
\newblock \bibinfo{title}{A short review of advances in the modelling of blood
  rheology and clot formation}.
\newblock {\it \bibinfo{journal}{Fluids}\/},  {\it \bibinfo{volume}{2}\/},
  \bibinfo{pages}{35}.
\bibitem[{Apostolidis et~al.(2015)Apostolidis, Armstrong \&
  Beris}]{apostolidis2015}
\bibinfo{author}{Apostolidis, A.~J.}, \bibinfo{author}{Armstrong, M.~J.}, \&
  \bibinfo{author}{Beris, A.~N.} (\bibinfo{year}{2015}).
\newblock \bibinfo{title}{Modeling of human blood rheology in transient shear
  flows}.
\newblock {\it \bibinfo{journal}{Journal of Rheology}\/},  {\it
  \bibinfo{volume}{59}\/}, \bibinfo{pages}{275--298}.
\bibitem[{Ballyk et~al.(1994)Ballyk, Steinman \& Ethier}]{ballyk1994}
\bibinfo{author}{Ballyk, P.}, \bibinfo{author}{Steinman, D.}, \&
  \bibinfo{author}{Ethier, C.} (\bibinfo{year}{1994}).
\newblock \bibinfo{title}{Simulation of non-newtonian blood flow in an
  end-to-side anastomosis}.
\newblock {\it \bibinfo{journal}{Biorheology}\/},  {\it
  \bibinfo{volume}{31}\/}, \bibinfo{pages}{565--586}.
\bibitem[{Bird et~al.(1987)Bird, Armstong \& Hassager}]{bird1987}
\bibinfo{author}{Bird, R.~B.}, \bibinfo{author}{Armstong, R.~C.}, \&
  \bibinfo{author}{Hassager, O.} (\bibinfo{year}{1987}).
\newblock {\it \bibinfo{title}{Dynamics of polymeric liquids {Volume 1: fluid
  mechanics}}\/}.
\newblock (\bibinfo{edition}{1st} ed.).
\newblock \bibinfo{publisher}{John Wiley \& Sons, Inc.}
\bibitem[{Carreau(1972)}]{carreau1972}
\bibinfo{author}{Carreau, P.~J.} (\bibinfo{year}{1972}).
\newblock \bibinfo{title}{Rheological {Equations} from {Molecular} {Network}
  {Theories}}.
\newblock {\it \bibinfo{journal}{Transactions of the Society of Rheology}\/},
  {\it \bibinfo{volume}{16}\/}, \bibinfo{pages}{99--127}.
\bibitem[{Cokelet et~al.(1963)Cokelet, Merrill, Gilliland, Shin, Britten \&
  Wells~Jr}]{cokelet1963}
\bibinfo{author}{Cokelet, G.~R.}, \bibinfo{author}{Merrill, E.~W.},
  \bibinfo{author}{Gilliland, E.~R.}, \bibinfo{author}{Shin, H.},
  \bibinfo{author}{Britten, A.}, \& \bibinfo{author}{Wells~Jr, R.~E.}
  (\bibinfo{year}{1963}).
\newblock \bibinfo{title}{The rheology of human blood-measurement near and at
  zero shear rate}.
\newblock {\it \bibinfo{journal}{Transactions of the Society of Rheology}\/},
  {\it \bibinfo{volume}{7}\/}, \bibinfo{pages}{303--317}.
\bibitem[{Cross(1965)}]{cross1965}
\bibinfo{author}{Cross, M.~M.} (\bibinfo{year}{1965}).
\newblock \bibinfo{title}{Rheology of non-{N}ewtonian fluids: a new flow
  equation for pseudoplastic systems.}
\newblock {\it \bibinfo{journal}{Journal of Colloid and Interface Science}\/},
  {\it \bibinfo{volume}{20}\/}, \bibinfo{pages}{417--437}.
\bibitem[{Doke(2016)}]{grabit}
\bibinfo{author}{Doke, J.} (\bibinfo{year}{2016}).
\newblock \bibinfo{title}{{GRABIT}}.
\newblock
  \bibinfo{howpublished}{\url{https://uk.mathworks.com/matlabcentral/fileexchange/7173-grabit}}.
\newblock \bibinfo{note}{Accessed: 2018-09-24}.
\bibitem[{Gijsen et~al.(1999)Gijsen, van~de Vosse \& Janssen}]{gijsen1999}
\bibinfo{author}{Gijsen, F. J.~H.}, \bibinfo{author}{van~de Vosse, F.~N.}, \&
  \bibinfo{author}{Janssen, J.~D.} (\bibinfo{year}{1999}).
\newblock \bibinfo{title}{The influence of the non-{N}ewtonian properties of
  blood on the flow in large arteries: steady flow in a carotid bifurcation
  model}.
\newblock {\it \bibinfo{journal}{Journal of Biomechanics}\/},  {\it
  \bibinfo{volume}{32}\/}, \bibinfo{pages}{601--608}.
\bibitem[{Huang et~al.(1973)Huang, King \& Copley}]{huang1973}
\bibinfo{author}{Huang, C.~R.}, \bibinfo{author}{King, R.~G.}, \&
  \bibinfo{author}{Copley, A.~L.} (\bibinfo{year}{1973}).
\newblock \bibinfo{title}{Rheogoniometric studies of whole human blood at shear
  rates down to 0.0009 sec-1}.
\newblock {\it \bibinfo{journal}{Biorheology}\/},  {\it
  \bibinfo{volume}{10}\/}, \bibinfo{pages}{23--28}.
\bibitem[{Kolbasov et~al.(2016)Kolbasov, Comiskey, Sahu, Sinha-Ray, Yarin,
  Sikarwar, Kim, Jubery \& Attinger}]{kolbasov2016}
\bibinfo{author}{Kolbasov, A.}, \bibinfo{author}{Comiskey, P.~M.},
  \bibinfo{author}{Sahu, R.~P.}, \bibinfo{author}{Sinha-Ray, S.},
  \bibinfo{author}{Yarin, A.~L.}, \bibinfo{author}{Sikarwar, B.~S.},
  \bibinfo{author}{Kim, S.}, \bibinfo{author}{Jubery, T.~Z.}, \&
  \bibinfo{author}{Attinger, D.} (\bibinfo{year}{2016}).
\newblock \bibinfo{title}{Blood rheology in shear and uniaxial elongation}.
\newblock {\it \bibinfo{journal}{Rheologica Acta}\/},  {\it
  \bibinfo{volume}{55}\/}, \bibinfo{pages}{901--908}.
\bibitem[{Leuprecht \& Perktold(2001)}]{leuprecht2001}
\bibinfo{author}{Leuprecht, A.}, \& \bibinfo{author}{Perktold, K.}
  (\bibinfo{year}{2001}).
\newblock \bibinfo{title}{Computer simulation of non-newtonian effects on blood
  flow in large arteries}.
\newblock {\it \bibinfo{journal}{Computer Methods in Biomechanics and
  Biomedical Engineering}\/},  {\it \bibinfo{volume}{4}\/},
  \bibinfo{pages}{149--163}.
\bibitem[{Merrill et~al.(1963)Merrill, Gilliland, Cokelet, Shin, Britten \&
  Wells~Jr}]{merrill1963}
\bibinfo{author}{Merrill, E.~W.}, \bibinfo{author}{Gilliland, E.~R.},
  \bibinfo{author}{Cokelet, G.}, \bibinfo{author}{Shin, H.},
  \bibinfo{author}{Britten, A.}, \& \bibinfo{author}{Wells~Jr, R.~E.}
  (\bibinfo{year}{1963}).
\newblock \bibinfo{title}{Rheology of human blood, near and at zero flow.
  effects of temperature and haematocrit level}.
\newblock {\it \bibinfo{journal}{Biophysical Journal}\/},  {\it
  \bibinfo{volume}{3}\/}, \bibinfo{pages}{199--213}.
\bibitem[{Ostwald(1925)}]{ostwald1925}
\bibinfo{author}{Ostwald, W.} (\bibinfo{year}{1925}).
\newblock \bibinfo{title}{Ueber die geschwindigkeitsfunktion der viskosit{\"a}t
  disperser systeme. i}.
\newblock {\it \bibinfo{journal}{Kolloid-Zeitschrift}\/},  {\it
  \bibinfo{volume}{36}\/}, \bibinfo{pages}{99--117}.
\bibitem[{Perktold \& Rappitsch(1995)}]{perktold1995}
\bibinfo{author}{Perktold, K.}, \& \bibinfo{author}{Rappitsch, G.}
  (\bibinfo{year}{1995}).
\newblock \bibinfo{title}{Computer simulation of local blood flow and vessel
  mechanics in a compliant carotid artery bifurcation model}.
\newblock {\it \bibinfo{journal}{Journal of Biomechanics}\/},  {\it
  \bibinfo{volume}{28}\/}, \bibinfo{pages}{845--856}.
\bibitem[{Skalak et~al.(1981)Skalak, Keller \& Secomb}]{skalak1981}
\bibinfo{author}{Skalak, R.}, \bibinfo{author}{Keller, S.~R.}, \&
  \bibinfo{author}{Secomb, T.~W.} (\bibinfo{year}{1981}).
\newblock \bibinfo{title}{Mechanics of blood flow}.
\newblock {\it \bibinfo{journal}{Journal of Biomechanical Engineering}\/},
  {\it \bibinfo{volume}{103}\/}, \bibinfo{pages}{102--115}.
\bibitem[{de~Waele(1923)}]{deWaele1923}
\bibinfo{author}{de~Waele, A.} (\bibinfo{year}{1923}).
\newblock \bibinfo{title}{Viscometry and plastometry}.
\newblock {\it \bibinfo{journal}{Journal of the Oil \& Colour Chemists'
  Association}\/},  {\it \bibinfo{volume}{6}\/}, \bibinfo{pages}{33--69}.
\bibitem[{Yasuda(1979)}]{yasuda1979}
\bibinfo{author}{Yasuda, K.} (\bibinfo{year}{1979}).
\newblock {\it \bibinfo{title}{Investigation of the analogies between
  viscometric and linear viscoelastic properties of polystyrene fluids}\/}.
\newblock \bibinfo{type}{Thesis} Massachusetts Institute of Technology.

\end{thebibliography}
\end{document}